\documentclass[
    final            
  ]
  {aipproc}
\layoutstyle{6x9}

\begin{document}
\title[Compact binaries with an accreting white dwarf in globular clusters]{Formation and evolution of compact binaries with an accreting white dwarf in globular clusters.}

\author{N.~Ivanova}{
  address={Department of Physics and Astronomy, 2145 Sheridan Rd, Evanston, IL 60208, USA}}
\author{F.~A.~Rasio}{
  address={Department of Physics and Astronomy, 2145 Sheridan Rd, Evanston, IL 60208, USA}}
\begin{abstract}
The population of compact binaries in dense stellar systems 
is affected strongly by frequent dynamical interactions
between stars and their interplay with the stellar evolution.
In this contribution, we consider these effects on  
binaries with a white dwarf accretor, in particular cataclysmic variables and
AM~CVns.
We examine which processes can successfully lead to the creation of such X-ray binaries.
Using numerical simulations, we identify predominant formation channels 
and predict the expected numbers of detectable systems.
We discuss also why the distribution of cataclysmic variables 
has a weaker dependence upon the cluster density than the distribution of 
quiescent low-mass X-ray binaries and why dwarf nova outbursts may not occur among
globular cluster cataclysmic variables.
\end{abstract}
\maketitle

\section{Introduction}
In the past few years,  substantial progress has been made in
optical identification of {\it Hubble Space Telescope} counterparts to {\it Chandra}
X-ray sources in several globular clusters (GCs).
Valuable information was obtained for the population of cataclysmic variables (CVs), 
chromospherically active binaries and quiescent low-mass X-ray binaries (qLMXBs) 
(Edmonds et al. 2004, Heinke et al. 2003b, Haggard et al.  2004).
For the first time we can compare populations of such binaries in
GCs and in the Galactic field, as well as calculate the rates of
formation of these binaries and their population characteristics.
In particular, as many as 22 CVs have been already identified in 47 Tuc, posing a number
of interesting questions: e.g., their ratio of X-ray flux to optical flux is higher than 
in field CVs  (Edmonds et al. 2003) and they do not 
show dwarf nova outbursts, which are common for field CVs (Shara et al. 1996). 
Also, it has been found that the specific incidence of harder X-ray sources 
(primarily CVs) depends more weakly on density than that of qLMXBs (Heinke et al. 2003a).

The present population of close binaries with a white dwarf (WD) accretor in GCs
has not necessarily evolved from  primordial binaries, 
since it can be influenced by dynamical encounters. 
In this contribution we study these processes by combining an advanced binary population code and
careful treatment of all dynamical interactions with a simplified dynamical cluster background model.
We first consider what mechanisms are likely to produce a CV or an AM~CVn and
then compare our predictions with the results of numerical simulations.
We also discuss how the formation of mass-transfer systems with a WD accretor 
is different from the formation of these with systems with a neutron star (NS) accretor.

\section{Binaries in a dense cluster}

There are several ways to destroy a primordial binary in a globular cluster.
If the binary is in a dense region, dynamical encounters play a significant role in its evolution.
For instance, there is a high probability  to ``ionize'' a soft binary
as a result of a dynamical encounter. 
In the case of a hard binary, destruction can happen through a physical collision during the encounter.
Additionally, a primordial binary can be destroyed through an evolutionary merger 
(most important for hard binaries) or following to a SN explosion. 
Overall, we find that, if a typical cluster initially had
as many as 100\% of its stars in primordial binaries,
the binary fraction at an age of 10-14 Gyr will be only $\le 10\%$ (Ivanova et al. 2005).

The typical formation scenario for CVs in the field (low density environment)
involves the common envelope (CE) evolution.  
The minimum initial orbital period for the progenitor binary is about 100 days
and the primary mass has to be smaller than 8 $M_\odot$.
In the core of a GC with core density $\rho_{\rm c} \sim 10^5$ pc$^{-3}$, 
a binary with such a period will experience a dynamical encounter before its primary 
leaves the main sequence.
The primordial channel for the CV formation could succeed therefore 
only if this binary entered the dense cluster core after the CE event.
The contribution of the primordial channel depends mainly on the cluster
half-mass relaxation time, which regulates how fast binaries will segregate into the central dense region.
The situation is even more striking for the formation of AM~CVn.
In this case, the standard formation channel requires the occurrence of two CE events,
and the primordial binary is expected to be even wider.

\subsection{Dynamical formation of CVs.}

There are two main types of dynamical encounters that lead to the dynamical formation of a
binary consisting of a main sequence star (MS) and a WD:
exchange interactions, and physical collisions between a red giant (RG) and a MS star. 
However, only a fraction of the dynamically formed MS-WD binary systems is capable of becoming a CV.
We will consider first the post-encounter binary evolution of a dynamically formed MS-WD binary
(to find out which post-encounter parameters favor CV formation), and
then determine which encounters give these specific parameters.

A close MS-WD binary looses its angular momentum through  magnetic braking (MB)
and  gravitational wave (GW) emission. 
The orbital shrinkage due to the synchronization of the MS star with the orbital motion  
can be neglected: less than a few percent of the total orbital angular momentum
is required to spin-up a MS.
A post-exchange hard binary has an average eccentricity  $e\sim 0.7$, following a thermal distribution,
and a post-collision binary has eccentricity  $e \ge 0.4$ (Lombardi et al. 2005).

\begin{figure}
  \includegraphics[height=.35\textheight]{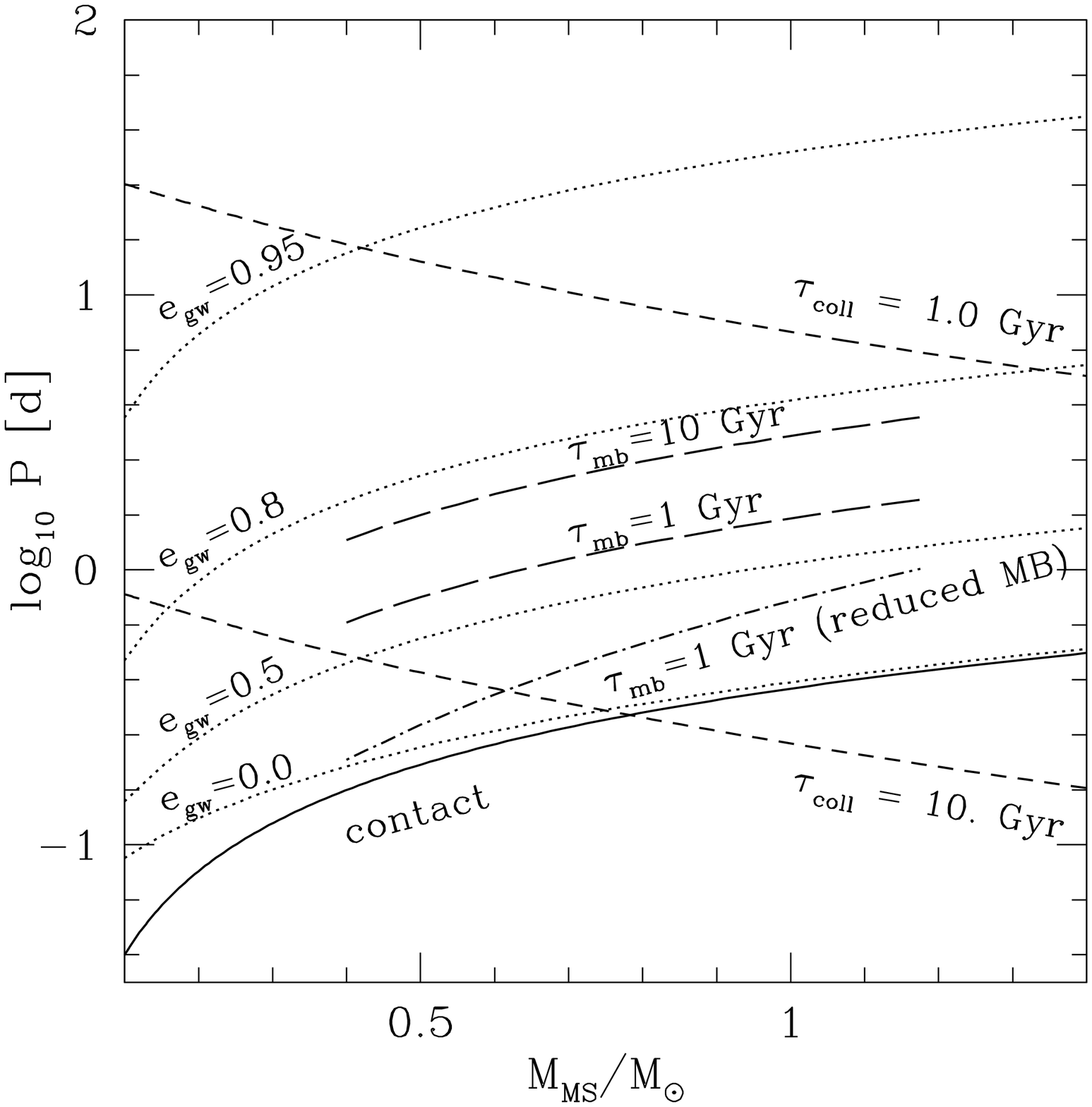}
  \includegraphics[height=.35\textheight]{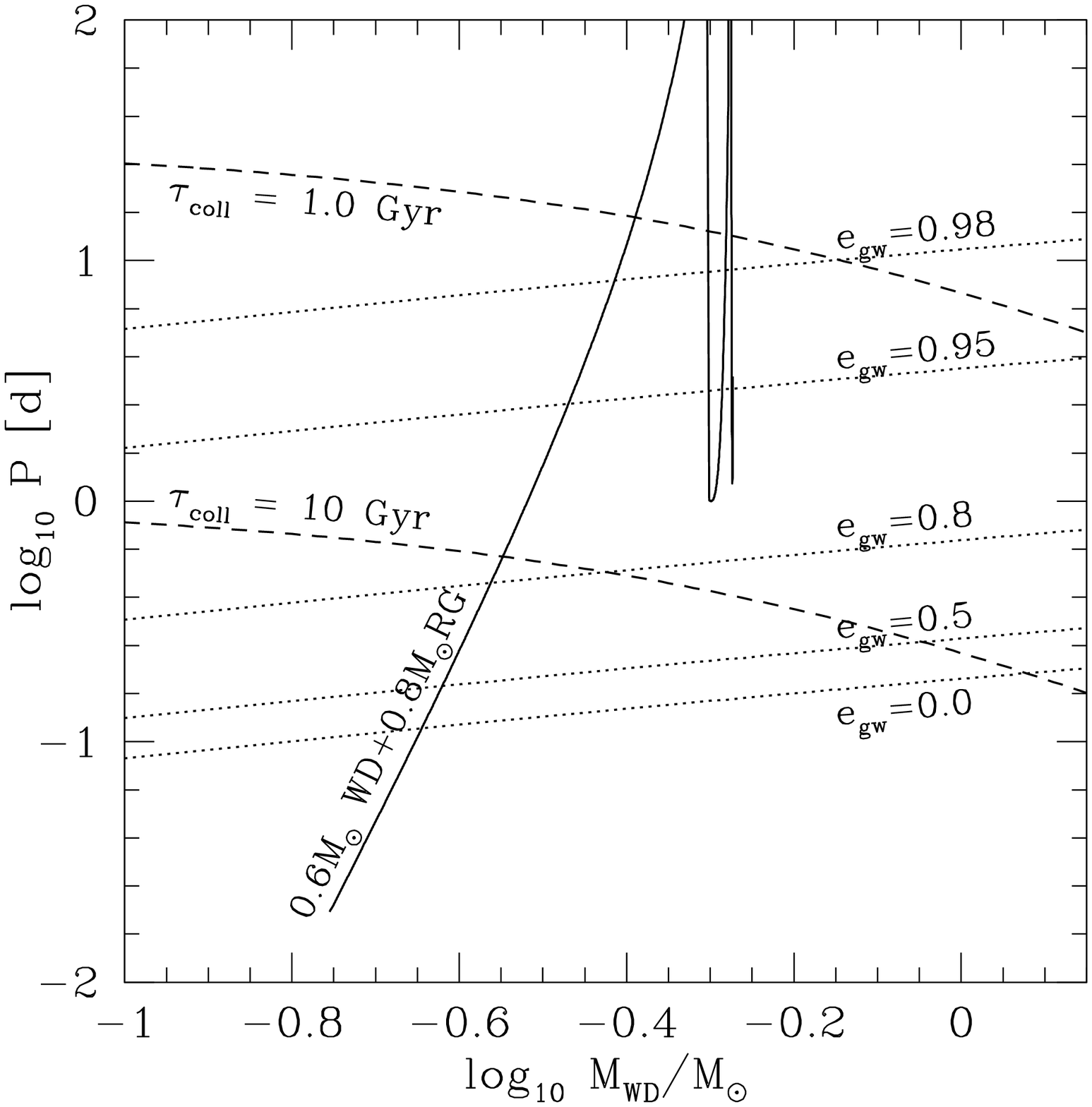}
  \caption{The fate of post-encounter binaries.
{\bf Left panel:} MS-WD binaries where the primary is a WD of 0.6 $M_\odot$.
$P$ is the post-encounter orbital period and $M_{\rm MS}$ is the mass of a MS secondary.
The short-dashed lines show the binary periods for collision times of 1 Gyr and 10 Gyr.
The dotted lines delineate the binaries that will shrink within 1 Gyr due to gravitational wave emission, 
for different post-exchange eccentricities. 
The long-dashed lines delineate the binaries that will shrink within 1~Gyr and 10~Gyr
for the standard magnetic braking prescription, and dash-dotted line -- for 
reduced magnetic braking.
Below the solid line the binary is in contact.
{\bf Right panel:} WD-WD binaries where the primary is a WD of 0.6 $M_\odot$.
$P$ is the post-encounter orbital period and $M_{\rm WD}$ is the mass of a WD secondary.
The dashed and dotted lines are the same as in the left panel.
The solid line shows binaries that can be formed through a collision between a WD of 0.6 $M_\odot$
and a RG of 0.8 $M_\odot$ ( $\alpha_{\rm CE} \lambda=1$).
}
\end{figure}

Depending on the MB prescription  -- the standard MB (Skumanich 1972, Pylyser \& Savonije 1988) 
or the reduced MB (Ivanova \& Taam 2003) -- in a dynamically formed eccentric binary with $e\approx 0.7$, 
MB is at most comparable to GW emission (see Fig.~1).
In order to become a CV within 1 Gyr, a binary should have a post-encounter
period $ \sim 1$ day  (the separation is then $\approx 10\, R_\odot$), or
$P \sim 10$ day and very high eccentricity, $e\ge0.96$.
It is not clear which post-exchange binaries will dominate in producing CVs.
At first glance, in the encounter with a hard binary, the post-exchange separation is 
comparable to the separation in the pre-encounter binary. 
Therefore, considering the collision time, the formation of  a wider binary through exchange
is more likely. On the other hand, it is rather unlikely to produce such a very high eccentricity
through an exchange.

In the case of a MS-WD binary with  $P \ge 10$ days and a moderate eccentricity $e\le 0.8$,
dynamical interactions that occur after  binary formation cannot harden such a binary significantly.
Let us consider a binary consisting of a 1 $M_\odot$ MS and a 0.6 $M_\odot$ WD.
Even if each hardening encounter could reduce the orbital separation by $ 50\%$, 
the hardening of this binary from 10 days to 1 day  
(at this period MB starts to be efficient) will take about 20 Gyr.
In contrast to hardening, another mechanism is important: {\em eccentricity pumping}.
The mean time between successive collisions $\tau_{\rm coll}\le1$~Gyr 
and therefore a binary can experience many encounters. 
If the acquired eccentricity $e\ge0.95$, 
the binary can shrink through GW emission even if its initial period is bigger than 10 days.

As the post-exchange separation in a binary that becomes
 a CV is comparable to the separation in the pre-encounter binary,
the pre-encounter binary has to be very tight.
It is not likely that, in the case of a tight binary (with moderate eccentricity),
the pre-encounter binary was a MS-MS binary or a MS-WD binary: 
an encounter with so close a binary will likely lead to a physical 
collision rather than to an exchange (Fregeau et al. 2004).
This restriction, together with $\tau_{\rm coll}\sim$~few Gyr for pre-collision binaries 
of $P\le 1$~day, predicts that most of the post-exchange CVs
can be formed in a three-body encounter where the pre-encounter binary had $P \ge  3$ days and
the post-encounter binary has high eccentricity $e\ge 0.8$ (or this eccentricity was
increased in subsequent encounters). No hardening is expected.

For binaries formed through MS-RG collisions, the binary separation can be estimated
using the standard CE prescription with $\alpha_{\rm CE} \lambda=1$ (Iben \& Livio 1993). 
The consideration of a RG inner structure through its evolution predicts that
only collision of a MS star with a RG with core mass $M_{\rm core}\le 0.3 M_\odot$  or with a giant at the
core helium burning stage with $M_{\rm core}\approx 0.6\, M_\odot$ will lead to the formation 
of a tight enough binary (see also Fig.~1, right panel).

A WD of 0.3 $M_\odot$ cannot be formed unless it evolved through a CE event or
a physical collision. A post-CE binary for this WD is very hard and has $\tau_{\rm coll}\ge 10$~Gyr, 
so the number of single $0.3 M_\odot$ WDs is negligible.
This restricts the exchange formation channel for CVs with a low-mass WD:
all CVs with a low mass WD companion must be formed either through a CE event
(in a primordial binary or in a dynamically formed binary with $P\sim 10-100$ days),
or as a result of a physical collision.

\subsection{Dynamical formation of AM CVn systems.}

Let us consider first a hard post-exchange WD-WD binary.
As we described previously, the typical eccentricity  is $e\sim0.7$, the separation is 
comparable to the pre-exchange separation; there is no post-exchange hardening.
The main difference with MS-WD binaries is that post-exchange WD-WD binary periods
that will allow a binary to evolve to mass transfer (MT) are several times smaller 
(see Fig.~1, right panel).
The collision time for pre-encounter binaries is about 10 Gyr, therefore making
the formation an AM CVn binary through three-body exchange (with subsequent 
GW decay) rather unlikely.

A variant for this channel is the dynamical formation of a relatively wide MS-WD 
binary that will subsequently evolve through CE.
The collisional time for these relatively wide MS-WD binaries is less than 1 Gyr and therefore
a significant fraction of such binaries may participate in some other encounter
(destruction, hardening or eccentricity pumping) before the CE occurs.

The second important channel is again a physical collision, in this case of a single WD
with a RG (see Fig.~1,  where we show possible outcomes of a such a collision). 

We therefore expect that only a post-CE system can become an AM~CVn, where the post-CE system could
be from a primordial binary, a post-collision binary, or a dynamically formed binary.

\section{Numerical results}

\begin{figure}
  \includegraphics[height=.35\textheight]{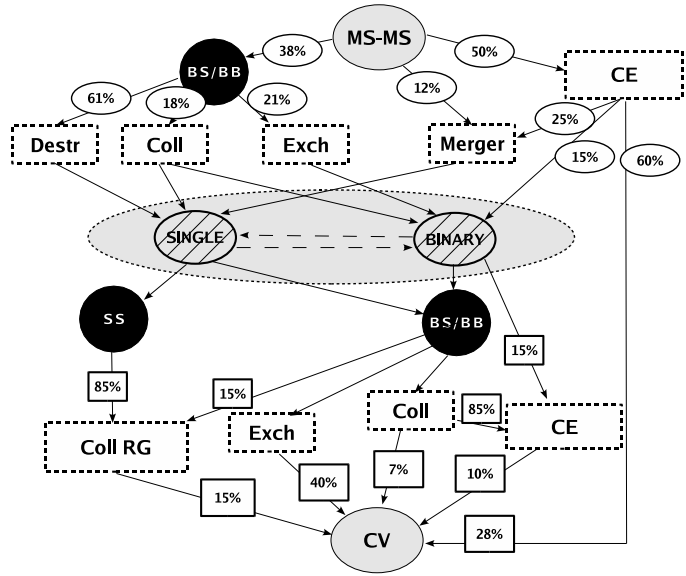}
  \caption{Main formation channels of CVs in a typical cluster.}
\end{figure}

To study stellar evolution in a dense system, we use a Monte Carlo approach which
couples one of most advanced binary population synthesis codes
currently available (Belczynski et al. 2002), a simple model for the cluster, and a small 
$N$-body integrator for
accurate treatment of all relevant dynamical interaction processes (Fregeau et al. 2004).
The complete description of the method is provided in Ivanova et al. 2004.
Here we treated physical collisions with a simple CE prescription
with post-collision $e=0$, though in a more realistic treatment the post-collision 
binary parameters should depend on the impact parameter (c.f. Lombardi et al. 2005).

We studied a ``typical'' cluster, starting with  $N=10^6$ stars and initial binary fraction 100\%. 
This high primordial binary fraction is  needed in order to match the observed binary fraction in GC cores today
(Ivanova 2005). The core density  $\rho_{\rm c}=10^{4.7} M_\odot \ {\rm pc}^{-3}$, one-dimensional velocity
dispersion $\sigma_1=10$ km/s,  escape speed from the cluster $v_{\rm esc}=43$ km/s,
and half-mass relaxation time $t_{\rm rh}=1$ Gyr. 
We also considered a 47~Tuc-type cluster, characterized by  $\rho_{\rm c}=10^{5.2} M_\odot \ {\rm pc}^{-3}$, 
$\sigma_1=11$ km/s, $v_{\rm esc}=60$ and $t_{\rm rh}=3$ Gyr. 
We adopt the broken power law IMF of Kroupa (2002) for single primaries, a flat mass-ratio
distribution for secondaries and the distribution of initial binary periods constant
in logarithm between contact and $10^7$~d.
The mass of the cluster at 11 Gyr is $\sim 2\times 10^5 M_\odot$.

In Fig.~2 we show the main formation channels for CVs that operate in a typical cluster at 11 Gyr.
The most important channel for CV formation is through an exchange
encounter -- it provides 40\% of CVs, and also 7\% of CVs are in binaries that experienced a merger
during the last three-body encounter;
15\% of CVs were formed as a result of a physical collision between a RG and a MS star;
in 10\% the CE occured in a dynamically formed binary; and
28\% of CVs are provided by the primordial channel. In total, the number of
post-CE systems and post-exchange systems is about the same.

Typical participants of a  successful physical collision (leading to  CV formation) 
are a  MS star of $0.3-0.9~M_\odot$ and 
a RG of about $1-1.7 M_\odot$  with a core around $0.3 M_\odot$ or 
He core burning giant with  core mass around $0.5 M_\odot$.
CVs formed this way are similar to post-CE CVs from primordial binaries,
the typical post-exchange period is about 0.2 day.

\begin{figure}
  \includegraphics[height=.35\textheight]{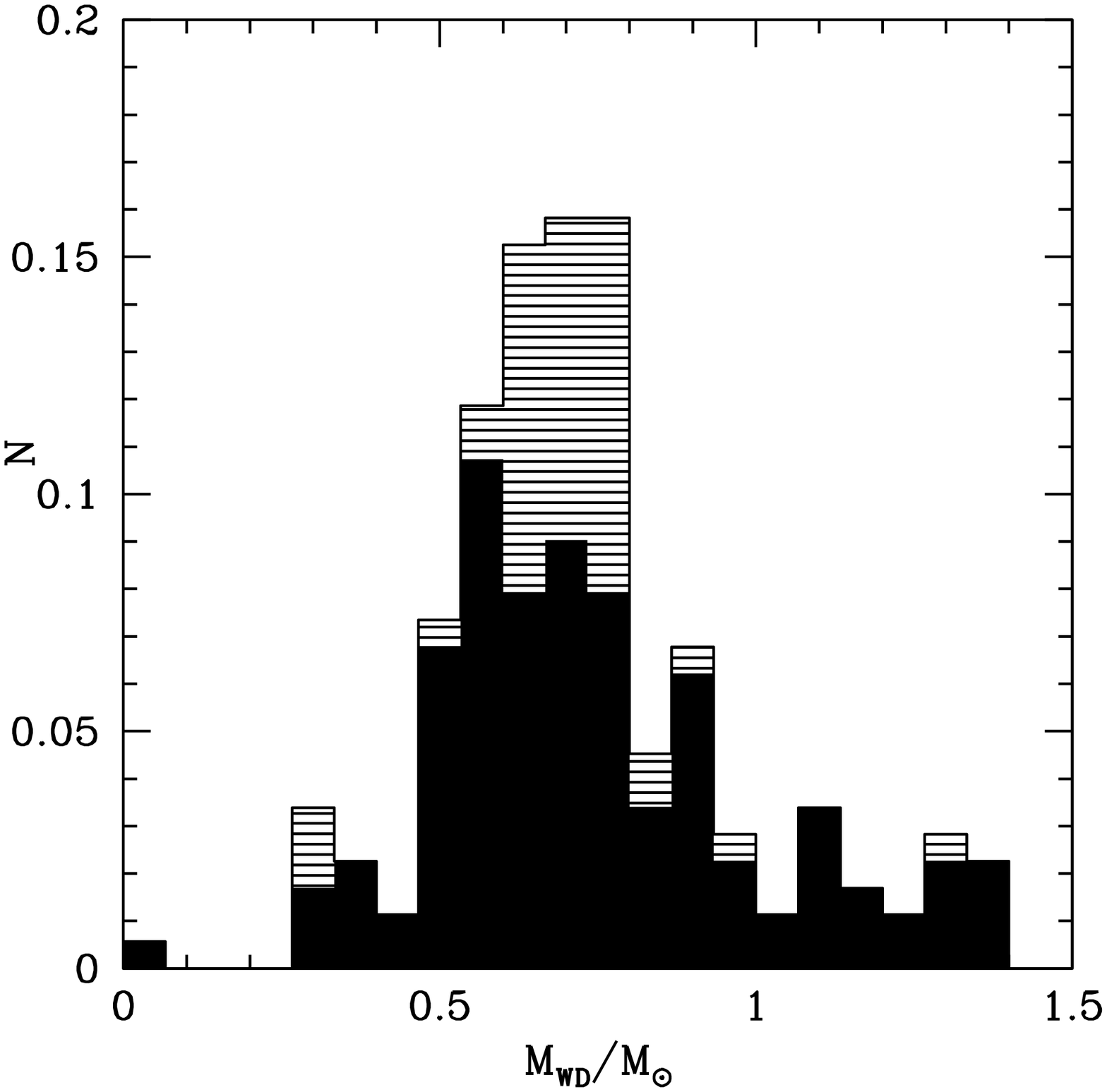}
  \includegraphics[height=.35\textheight]{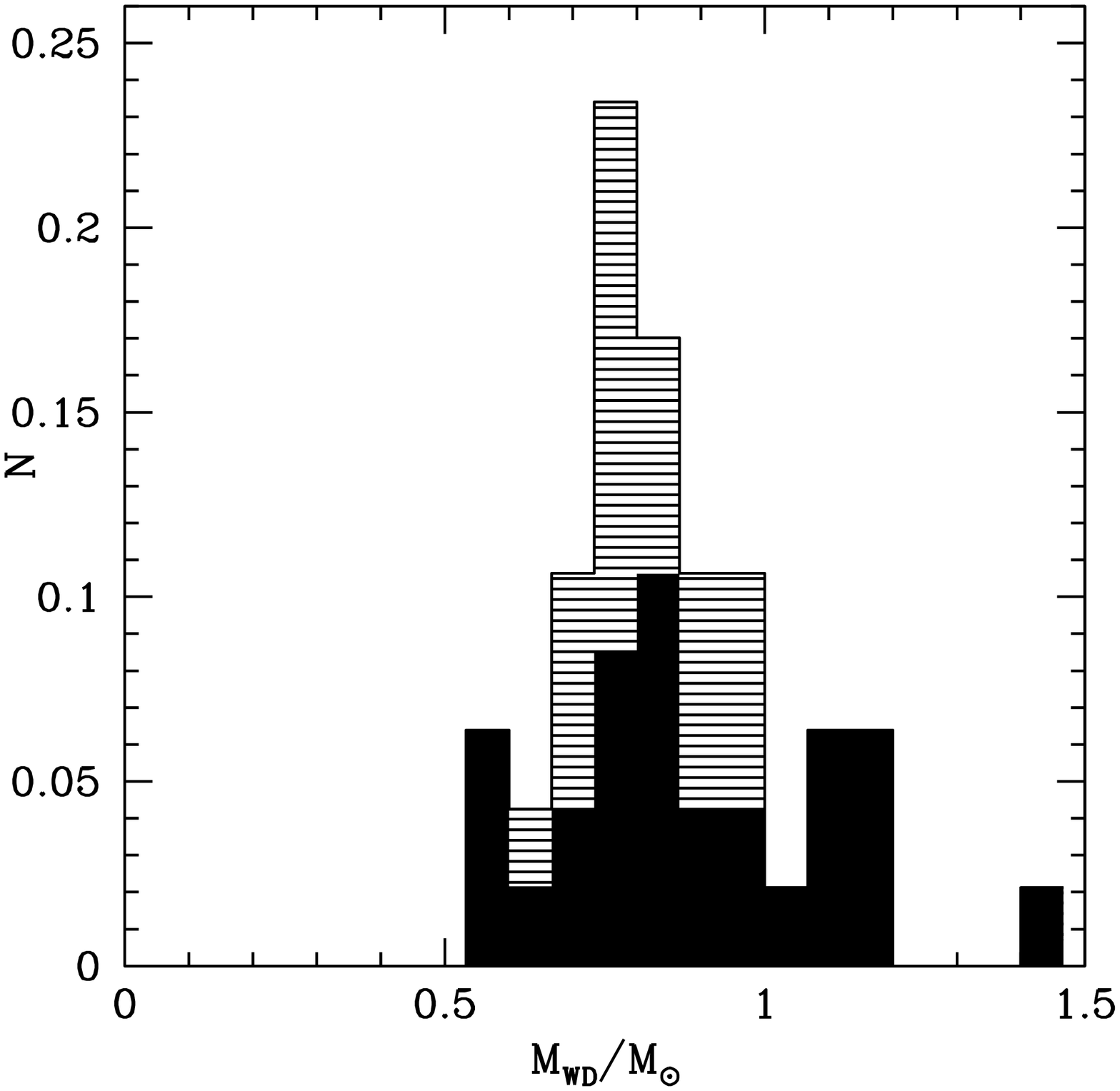}
  \caption{CVs (left) and AM~CVns (right) in a cluster core: the accreting WD-mass distribution. 
The solid filled area corresponds to dynamically formed binaries; the hatched area to systems 
formed directly from primordial binaries.}
\label{hist}
\end{figure}

For CVs formed through an exchange, there are two types of successful three-body encounters:
(a) a single, relatively heavy WD of about  $0.7-1.4 M_\odot$ and a MS-MS binary with a total mass usually 
$\le 1 M_\odot$;
(b) a single relatively massive MS star of around the turn-off mass and a MS-WD or WD-WD binary.
In most cases the WD-MS or WD-WD binary was not a primordial binary, but a dynamically formed binary.
The number of successful encounters between MS star and WD-WD binary is relatively small, and no 
successful four-body encounter occurred.
A post-exchange binary typically has a WD heavier than a post-CE binary.
As a result, the distribution of WD masses in CVs is more populated at the high mass end (see Fig.~\ref{hist})
compared to the field population, which has two very well distinguished peaks 
at $\sim 0.4$ and $\sim 0.5$ $M_\odot$ and exponentially decreases with WD mass (Willems, priv. comm.).
As expected, most CVs with a heavy WD were formed dynamically.

\begin{figure}
  \includegraphics[height=.35\textheight]{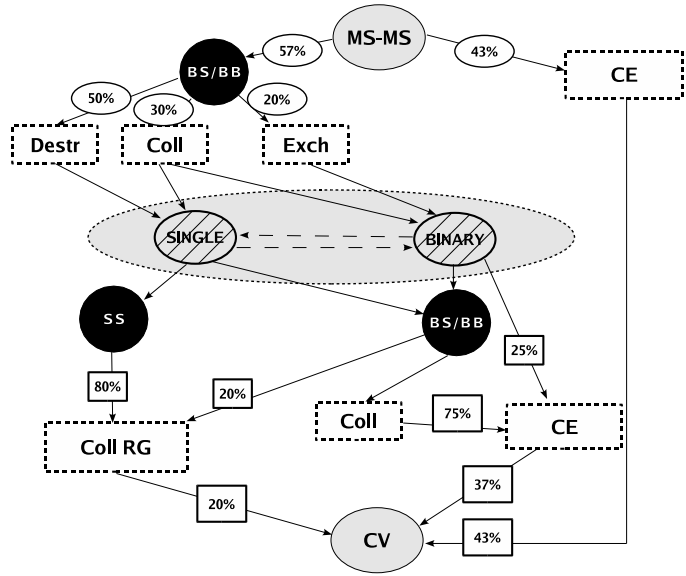}
  \caption{Main formation channels of AM~CVns in a typical cluster.}
\end{figure}

In Fig.~4 we show the main formation channels for AM~CVns that occur in a typical cluster.
As predicted, the main difference with CV formation is that there are only post-CE channels.
The role of physical collisions for AM~CVns  is slightly more important than for CVs, 
although this is only for the relative fraction.
The  number of  CVs formed by physical collisions is about 3 times higher than
the number  of AM~CVns formed by physical collisions. 
At the moment of a physical collision, in 2/3 of cases, the participants are a RG of $0.9-1.4 M_\odot$
and a relatively massive MS star with well developed He core.
The AM~CVns we see are the result of MT that started when the donor was at the late MS stage,
stripped down to the He core (the mass of the He donor in this system is typically about $0.01 M_\odot$),
or the system has passed through another CE event.
About a third of physical collisions occur between a RG and heavy WD.
Overall, this channel provides AM~CVns with accretor masses at present (11 Gyr) from $0.55 M_\odot$
to $1.4 M_\odot$, compared to the primordial channel where masses of accretors are mainly distributed between
$0.55 M_\odot$ and $0.9 M_\odot$ (Fig.~3; see also Nelemans et al. 2001).

The exchange formation channel for AM~CVns involves an exchange encounter and a CE event.
A single WD acquires a relatively massive companion that becomes later a RG or,
with a  0.6-0.8 $M_\odot$ MS star, a second encounter occurs, with either a physical 
collision between MS stars or exchange with a more massive MS.

The total numbers of CVs and AM~CVns present in a typical cluster are not very different from these in field population ---
they are comparable (per whole cluster population) and only 2-3 times larger in the core than in the field (per unit mass). 
Dynamical formation is responsible only for 60\%-70\% of CVs
in the core, but there is also a fraction of CVs that never entered the core.
The GC core density variation therefore does not play a significant role, in contrast to the 
case of NS binaries, where almost systems were formed dynamically (Ivanova et al. 2004) and the numbers have
a direct strong dependence on the cluster collision rate (Pooley et al. 2003).
We expect to have about 1 CV per $200-300\,M_\odot$ in the core of a typical cluster and
about 1 CV per $400-600\,M_\odot$ in a 47 Tuc type cluster (about 100 CVs in the core of 47 Tuc, in quite reasonable 
agreement with observations, Edmonds et al. 2003).
The number of AM CVns is typically 3-4 times smaller than the number of CVs.

One of the remaining questions is why CVs in GCs do not show nova outbursts.
There are many differences between the binary properties of GC and field CV populations.
From our simulations, we see that an accreting WD in a GC CV is typically more massive than in field,
and the MS donor is less massive.
A CV therefore is characterized by a smaller MT rate, $\dot M \sim10^{-11}\, M_\odot {\rm yr^{-1}}$, which is much
smaller than the typical value for dwarf novae ($\sim10^{-9}\, M_\odot\, {\rm yr^{-1}}$).
Nevertheless, we should mention this this value is large enough to provide $L_{X}\sim 10^{33} {\rm ergs\ s^{-1}}$,
which is about the maximum observed luminosity for CVs in, e.g., 47 Tuc.
Small MT rates also possibly explain why the optical fluxes from these objects are unexpectedly low in UV (Edmonds et al. 2003).
According to the disk instability model which is currently used to describe dwarf novae cycles
(Frank, King \& Raine 2002), the condition for instability to occur is described in terms of a critical MT rate.
For instance, there is a critical accretion rate below which a CV disk is cold and stable.
The accreting WD did not necessarily evolve through CE. 
This possibly results in a higher magnetic field WD.
A higher than usual magnetic field could also help suppress disk instabilities even further.

\begin{theacknowledgments}

We thank B.\ Willems and C. Heinke for helpful discussions. 
NI was supported by a {\em Chandra Theory} grant, FAR acknowledges support from NASA ATP Grant NAG5-12044.

\end{theacknowledgments}

\end{document}